\documentclass[preprint,showpacs]{revtex4} %,preprintnumbers,amsmath,amssymb]{revtex4}

\usepackage{graphicx}

\usepackage{dcolumn}% Align table columns on decimal point
\usepackage{bm}% bold math

\begin{document}

\title{Femtosecond Photo-Induced Multiphoton Analog Computation for Symmetry-Based Pattern Classification}
%\title{Analog Quantum Computation for Symmetry-Based Pattern Classification}

\author{Lev Chuntonov, Leonid Rybak, Andrey Gandman, and Zohar Amitay}
\email{amitayz@tx.technion.ac.il} %
\affiliation{Schulich Faculty of Chemistry, Technion - Israel
Institute of Technology, Haifa 32000, Israel}

\begin{abstract}
Multiphoton femtosecond coherent control of is used for implementing innovative photo-induced analog coherent computation that generally might
be a basis for future "smart hardware". The specific implemented computational task the classification of an unknown sequence into one of the
three groups: (i) a constant sequence that is composed of identical numbers, (ii) a sequence that is antisymmetric around a given point, or
(iii) neither. The input sequence is encoded into the spectral phases of a broadband femtosecond pulse and the computational task is being
carried out by the multiphoton nonlinear response of the irradiated physical system. Here, it is the simultaneous coherent two- and three-photon
absorption in atomic sodium (Na). The corresponding computational resources are the manifold of initial-to-final multiphoton excitation pathways
photo-induced by the broad spectrum of the irradiating femtosecond pulse. The answer is obtained by measuring only two observables, which are
two state-populations excited via the two- and three-photon absorption processes. Hence, the computational task is accomplished in a constant
number of operations irrespective of the sequence length. As such, the presented scheme, where the femtosecond pulse serves as a query and the
irradiated physical system serves as an oracle, provides a sufficient gain in the computational complexity.
\end{abstract}

%\begin{keywords}
%coherent control; intermediate-field regime; pulse shaping; sodium
%atom; antisymmetric phase patterns
%\end{keywords}


\maketitle

% main text

%\section{}

Analog quantum computation (AQC) utilizes the similarity between
the formal mathematical description of the physical phenomena
and the formulation of the problem being solved.
The information is processed by exploiting the physical phenomena directly:
it is the system evolution that processes the information and
performs a computational task \cite{cit:feynman, cit:lloyd-perspective, cit:brockett-sort}.
AQC is aimed to solve otherwise intractable problems, once the
similarity is found between the mutual relations of the problem
parameters and the corresponding physical quantities of the system
\cite{cit:brockett-sort, cit:farhi-adiabatic-comp, cit:zak-1,
cit:zak-2, cit:amemiya-1, cit:amemiya-2,
cit:weigert-diag-hermitian-matrix,
cit:georgoudas-earthquake-cel-aut, cit:girard-factor}.
Here we exploit the femtosecond coherent control
\cite{tannor_kosloff_rice_coh_cont,
shapiro_brumer_coh_cont_book, warren_rabitz_coh_cont,
rabitz_vivie_motzkus_kompa_coh_cont, dantus_exp_review-1,
dantus_exp_review-2} of multiphoton absorption to perform
a specific computational task:
classification of the unknown sequence in to the three groups:
a sequence composed of the identical components, a sequence that
is antisymmetric about a specific point, or neither.
The concept of using the femtosecond pulse as a query and an excited system
as an oracle provides a sufficient
gain in the computational complexity.
The experimentally demonstrated approach requires only
two measurements to accomplish the computational task
irrespectively of the sequence length.

Following from their coherence over a broadband spectrum,
femtosecond pulses are used as a unique tool for coherent control
and coherent spectroscopy of matter \cite{dantus_exp_review-1,
dantus_exp_review-2, silberberg_2ph_nonres1, silberberg_2ph_nonres2,
amitay_3ph_2plus1-1, amitay_3ph_2plus1-2, amitay_multi-channel}.
The broadband spectrum
provides a variety of photoinduced pathways that are used for the multiphoton
excitation.
All of the such pathways interfere between them and contribute to the finally
excited amplitude: constructive interference enhances the excited amplitude,
while destructive interference suppresses it.
The recently developed pulse-shaping techniques
\cite{pulse_shaping-weiner, pulse_shaping-gerber-1,
pulse_shaping-gerber-2} enables access to the parameters of
different frequency components of the pulse independently.
Among the parameters that
could be adjusted are phase, amplitude, and
polarization of the frequency components.
Control over these parameters is directly translated to the control over the
interference between the photoinduced pathways and, hence, to the
control over the excited amplitude.
%
%In the present work we utilize the phase-control .

The basic idea of the presented here experiments is to use the interference
pathways of the multiphoton excitation as the computational and information processing tool.
The unknown information to be processed is translated into the transformation
$\hat{U}$=exp$\left[i\Phi(\omega)\right]$
which is applied to the electric field of the excitation femtosecond
pulse with spectral amplitude $\left|E(\omega)\right|$.
This transformation results in a shaped pulse $\left|E(\omega)\right|$exp$\left[i\Phi(\omega)\right]$,
where $\Phi(\omega)$ is a spectral phase of frequency $\omega$.
The phase associated with each multiphoton pathway is equal to the sum of the
phases of all the involved photons.
All the possible combinations of the excitation pathways provided by the shaped pulse
interfere together during the excitation, i.e. evaluated at a single event,
as opposed to the classical information processing, where each information
component is evaluated separately.
After the excitation, the resulted excited amplitude contains the information
on the nature of the interference.
Relying on the knowledge of the light-matter interaction mechanism,
measurement of the excited amplitude can characterize the information originally used to shape the
excitation pulse.
Different physical systems respond in a different way to the excitation with a given pulse shape.
This is why the system used for the processing should be carefully chosen to suit the
specific computational task.
It is a general and very important attribute of AQC -- its
inherent problem-specific nature: one should always find a physical system
with quantities which mutual relation corresponds to these of the
problem to solve.

The computational task that we consider is a classification of the unknown
sequence in to the three groups:
a sequence composed of the identical components, which is referred as
{\it{constant}} sequence, a sequence that is antisymmetric about a
specific point - {\it{antisymmetric}} sequence, or neither, referred
as {\it{random}}.
For that purpose we encode the subjected sequences into the
phase patterns $\Phi_{const}$, $\Phi_{antisym}$, and $\Phi_{random}$ of the broadband spectrum.
Each component of the sequence is translated into the phase magnitude of the corresponding frequency
using the pulse shaping technique.
The physical process that suits the computational needs is the two-channel
femtosecond multiphoton absorption.
In the proposed scenario the amplitudes of the excitation channels characterize the
type of the sequence encoded into the transformation $\hat{U}$.
The different response by the different channels to the applied transformation
is a key property of the system that solves the presented classification task.

One of the excitation channels exploited for our computational task
is a non-resonant two-photon absorption channel
\cite{silberberg_2ph_nonres1, silberberg_2ph_nonres2}.
The excited amplitude in this case is proportional to
$A^{(2)}(\Omega)=\int^{+\infty}_{-\infty}E(\Omega/2-\alpha)E(\Omega/2+\alpha)d\alpha$.
It interferes all the possible two-photon pathways
with transition frequency $\Omega$,
each such pathway is composed of two photons with frequencies
$\Omega/2-\alpha$ and $\Omega/2+\alpha$, i.e., two frequencies located symmetrically
around $\Omega/2$.
The phase associated with each pathway is $\Phi(\Omega/2-\alpha)+\Phi(\Omega/2+\alpha)$.
For the transform-limited (TL) pulse, which is the shortest pulse for a given spectrum,
$\Phi(\omega)=0$ and all the excitation pathways interfere constructively.
The same applies to the transformations corresponding to the constant sequences
$\hat{U}^{const}$ with $\Phi(\omega)$=const as they add only a global phase factor,
identical to all the interfering pathways.
The two-photon non-resonant channel is also invariant to the transformations corresponding to
the antisymmetric sequences $\hat{U}^{antisym}$, while the point of the inversion corresponds to
half the two-photon transition frequency $\Omega/2$.
For that case
$\Phi(\Omega/2-\alpha)=-\Phi(\Omega/2+\alpha)$ holds, and all the available
pathways obtain zero relative phase. Hence, the application of $\hat{U}^{antisym}$
implies constructive interference between all the pathways.
On the other hand, along any other possible channel without special sensitivity to the symmetry of the
transformation, the induced excitation pathways interfere destructively.
Such channel is not invariant to $\hat{U}^{antisym}$, but to $\hat{U}^{const}$.

Combining all the described here together, we build a truth table for the sequence classification:
constant sequences induce fully constructive interference along both channels and they are
associated with a maximal amplitude of both excitation channels, which is actually the same
as for the TL pulse.
The antisymmetric sequences are associated with fully constructive interference along the two-photon
non-resonant channel, and with destructive interference along the channel without any symmetrical
properties.
The random sequences induce destructive interference along both channels and the excited amplitude
by the corresponding shaped pulses is sufficiently less then by the TL pulse.
To demonstrate this idea experimentally, we have chosen atomic two-channel system, which includes non-resonant
two-photon absorption channel coherently incorporated within a resonant-mediated (2+1) three
photon absorption channel.

\begin{figure} %[htbp]
\includegraphics[width=17cm]{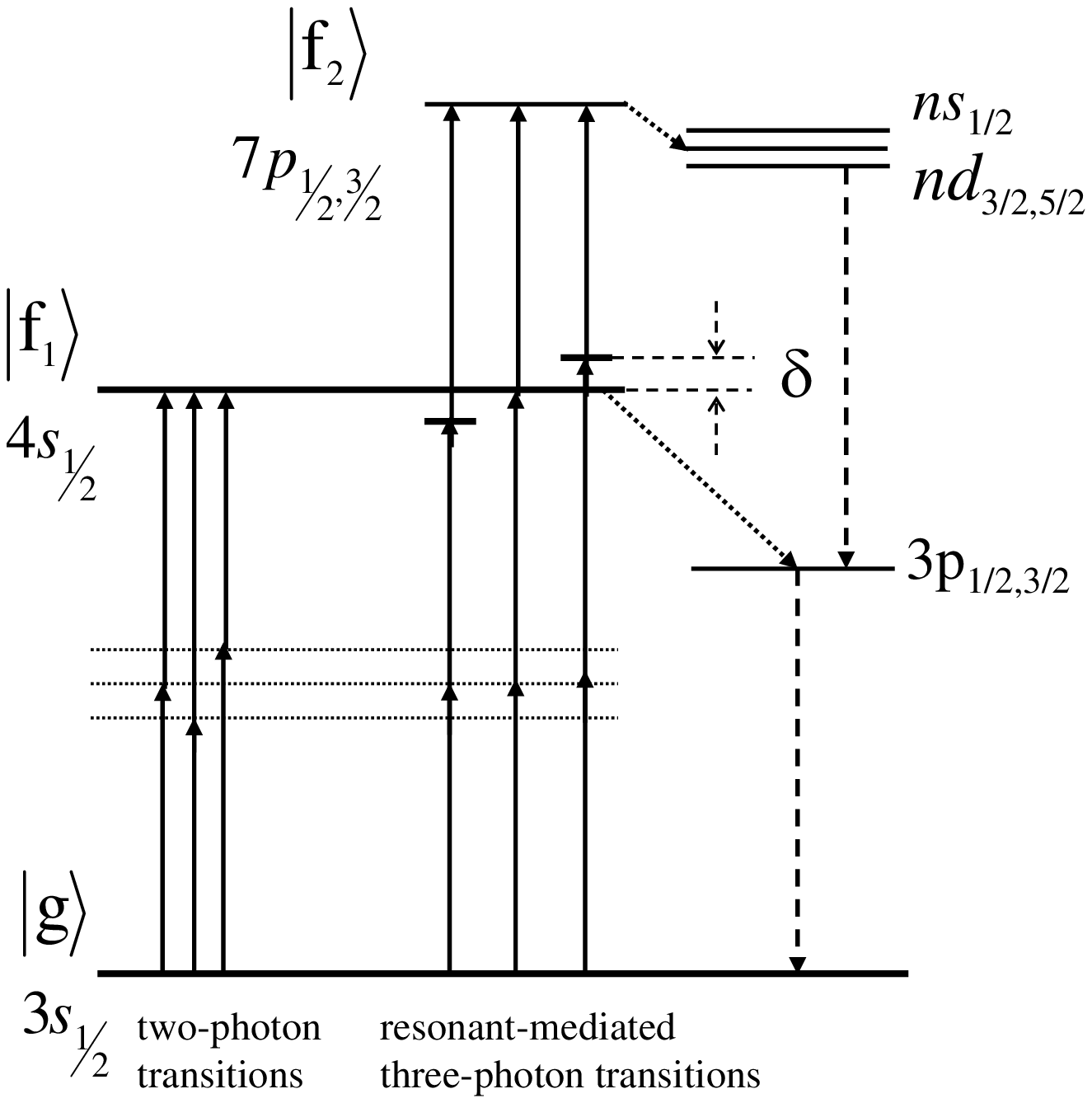}     %[scale=1.6]
\vspace*{-0.7cm} \caption{The two-channel femtosecond excitation
scheme of Na, including a non-resonant two-photon absorption channel
from $\left| g \right\rangle$$\equiv$$3s$ to $\left| f_1
\right\rangle$$\equiv$$4s$ and a resonance-mediated (2+1)
three-photon absorption channel from $\left| g
\right\rangle$$\equiv$$3s$ to $\left| f_2 \right\rangle$$\equiv$$7p$
via $\left| f_1 \right\rangle$$\equiv$$4s$.
Shown in solid lines are examples of two- and three-photon pathways.
The latter are either on resonance or near resonance with $\left|
f_1 \right\rangle$ (with detuning $\delta$).
The decay to the lower excited states is shown by dotted
lines, the measured radiative decay indicating the population of the
7p and 4s states is shown by dashed lines.
} \label{fig_1}
\end{figure}

The two-channel excitation considered here is shown schematically in
Fig.~\ref{fig_1}. It involves an initial ground state
$\left|g\right>$ and two excited states $\left|f_{1}\right>$ and
$\left|f_{2}\right>$.
The $\left|g\right>$ and $\left|f_{1}\right>$ states are coupled by
a non-resonant two-photon coupling provided by a manifold of states
$\left|n\right>$
that are far from resonance, while the $\left|f_{1}\right\rangle$ and
$\left|f_{2}\right\rangle$ states
are coupled resonantly by one-photon coupling.
Hence, when irradiated with a weak (shaped) femtosecond pulse of
temporal electric field $\varepsilon(t)$, two multiphoton absorption
channels are induced simultaneously:
a non-resonant two-photon absorption channel
from $\left|g\right>$ to $\left|f_{1}\right>$ is induced
simultaneously with
a resonance-mediated (2+1) three-photon absorption channel
from $\left|g\right>$ to $\left|f_{2}\right>$ with
$\left|f_{1}\right>$ as an intermediate state.
The amplitude of the latter channel is determined by
both intra-group and inter-group interferences corresponding to on-resonant and
near-resonant with $\left|f_{1}\right>$ state three-photon excitation pathways respectively.
The pathways induced by the pulse shaped with $\hat{U}^{antisym}$ interfere destructively
along this channel.
Although the TL pulse and, hence, the shaped pulses corresponding to $\hat{U}^{const}$
do not induce the maximal constructive interference for the (2+1)
resonance-mediated channel, the pulse should be shaped in a very special way
to enhance the amplitude excited by the TL pulse \cite{amitay_3ph_2plus1-1, amitay_3ph_2plus1-2, amitay_multi-channel}.
This is not the case of the present experiment.

Alternatively, another two-photon non-resonant channel can be chosen for that purpose
with different transition frequency $\Omega_{ch_1}\ne\Omega_{ch_2}$.
Obviously, while the channel 1 is sensitive to the sequences antisymmetric about $\Omega_{ch_1}/2$,
the channel 2 is sensitive to the antisymmetric sequence with the inversion point at $\Omega_{ch_2}/2$.
The corresponding physical system is a second harmonic crystal, with each frequency
of the produced harmonics being a different excitation channel measured separately.
The choice of such a crystal provides a wide tunability upon the point of antisymmetric inversion
of the sequences to be checked, while the sensitivity to this specific point
(the precision of the classification) in this case is set by the resolution of the detection system.
In the atomic case, this sensitivity is extremely high due to the narrow width
of the atomic lines.

Suppose now that unknown sequence $f(x)$ is optically encoded into
the phase pattern $\Phi(\omega)$ of the pulse using a pulse shaping
technique.
We check whether there is an inversion point $x_{0}$ corresponding to the
frequency of the non-resonant two-photon transition $\Omega/2=\omega_{f_{1},g}/2$.
The processing of the information is done upon the excitation of the atomic system with the shaped pulse
from its ground state $\left|g\right>$ to the excited states $\left|f_{1}\right>$ and $\left|f_{2}\right>$.
The population of the excited states $P_{f_{1}}$ and $P_{f_{2}}$
is used as a readout channel and compared with the population excited by the TL pulse.
If $P_{f_{1},f(x)}=P_{f_{1},TL}$ and $P_{f_{2},f(x)}=P_{f_{2},TL}$, the pulse spectral phase is a constant
pattern which represents encoded sequence $f(x)=const$.
If $P_{f_{1},f(x)}=P_{f_{1},TL}$, but $P_{f_{2},f(x)}\ne P_{f_{2},TL}$,
the pulse spectral phase is antisymmetric about $\omega_{fg}$ the encoded sequence satisfies
$f(x_{0}-\alpha)=-f(x_{0}+\alpha)$.
If $P_{f_{1},f(x)}\ne P_{f_{1},TL}$ and $P_{f_{2},f(x)}\ne P_{f_{2},TL}$,
the corresponding phase pattern is neither
antisymmetric nor constant and is referred here as random.
Overall, the computational task is performed by comparing the
outcome of the two specific measurements only, irrespectively of the
length of sequence to be classified.

The physical model system of the study is the sodium (Na) atom
\cite{NIST}, with the $3s$ ground state as $\left|g\right\rangle$,
the $4s$ state as $\left|f_1\right\rangle$, and the $7p$ state as
$\left|f_2\right\rangle$ (see Fig.~\ref{fig_1}).
The transition frequency $\omega_{f_1,g} \equiv \omega_{4s,3s} =
25740$~cm$^{-1}$ corresponds to two 777-nm photons and the
transition frequency $\omega_{f_2,f_1} \equiv \omega_{7p,4s} =
12801$~cm$^{-1}$ corresponds to a 781.2-nm photon.
The $3s$-$4s$ non-resonant two-photon coupling originates from the
manifold of $p$-states, particularly from the $3p$ state
[$\omega_{3p,3s}$$\sim$16978~cm$^{-1}$ (589~nm)].
The sodium is irradiated with phase-shaped linearly-polarized
femtosecond pulses having a Gaussian intensity spectrum centered
around 780~nm (12821~cm$^{-1}$) with 5.8-nm (95-cm$^{-1}$) bandwidth
($\sim$180-fs TL duration).

Experimentally, a sodium vapor in a heated cell is irradiated with
such laser pulses, after they undergo shaping in an optical setup
incorporating a pixelated liquid-crystal spatial light phase
modulator \cite{pulse_shaping-weiner}. The effective spectral
shaping resolution is $\delta\omega_{shaping}$=2.05~cm$^{-1}$
(0.125~nm) per pixel.
The peak intensity of the TL pulse is below 10$^{9}$~W/cm$^{2}$. %corresponding to the weak-field regime.
Following the interaction with a pulse, the population excited to
the $4s$ state radiatively decays to the lower $3p$ state, which
then decays to the $3s$ ground state. The $3p$-$3s$ fluorescence
serves as the relative measure for the total $4s$ population
$P_{f_1} \equiv P_{4s}$.
The population excited to~the $7p$ state undergoes radiative and
collisional decay to lower excited states, including the $4d$, $5d$,
$6d$, and $6s$ states.
The fluorescence emitted in their decay to the $3p$ state %(460-580-nm spectral range)
serves as the relative measure for the total $7p$ population
$P_{f_2} \equiv P_{7p}$.
The fluorescence is measured %optically measured %at 90$^{\circ}$ to the beam propagation direction
using a spectrometer coupled to a time-gated camera system.
The $3p$-$3s$ fluorescence part originating from the $4s$ state is
discriminated from the part originating from the $7p$ state
by using a proper %temporal
detection gate width, utilizing the different~time scales of the
$4s$-to-$3p$ and $7p$-to-$3p$ decays.

\begin{figure} %[htbp]
\includegraphics[width=17cm]{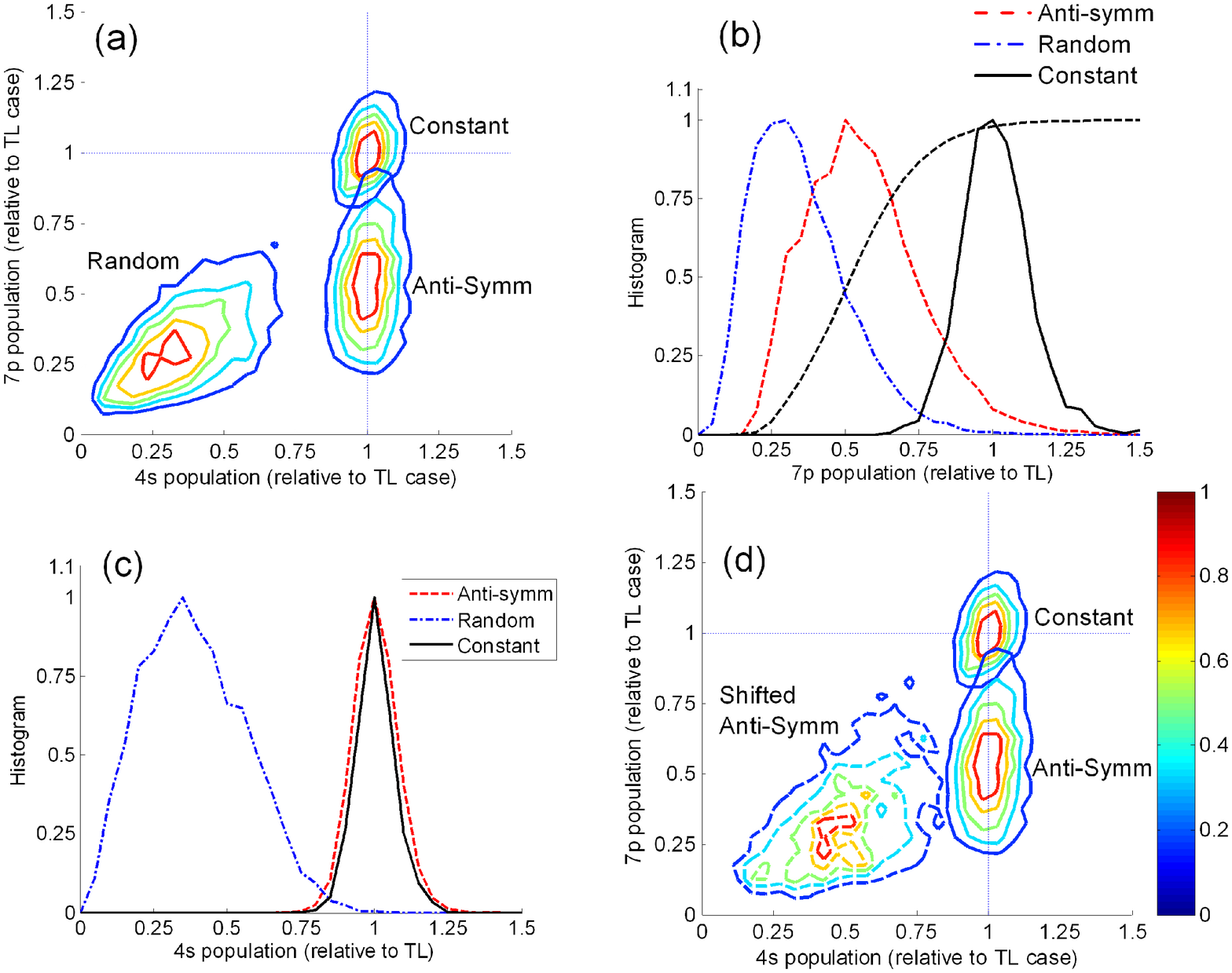}     %[scale=1.6]
\vspace*{-0.7cm} \caption{Experimental results for the 20,000 sequences encoded to the phase patterns of the shaped pulses. Each pattern has 26
bins with the values randomly generated within the interval of [$-\pi,\pi$]. Half of the sequences were urged to be antisymmetric about
$\omega_{4s,3s}$.
Panel (a): Presented is two-dimensional
histogram of the normalized to the unshaped (TL) pulse populations
of 4s and 7p states. Three regions corresponding to the
constant, antisymmetric, and random sequences are clearly
distinguished.
Panel (b): Projections of the panel (a) histogram on the axis of
7p population. Dashed black line indicates the cumulative sum of antisymmetric
sequences histogram. At $(x=1)$ it obtains the value of 0.96.
Panel (c): Projections of the panel (a) histogram on the 4s population axis.
Panel (d): Two-dimensional histogram  of antisymmetric sequences shifted by
$\Delta=\delta\omega_{shaping}$ from the original center of antisymmetry
$\omega_{4s,3s}/2$. The resulted sufficient shift of the distribution
indicates the high sensitivity of the proposed method.
%See the text for detailed description.}
} \label{fig_2}
\end{figure}

Fig.~\ref{fig_2} presents experimental results measured for the
randomly generated 20,000 sequences encoded into the phase patterns.
Half of the sequences are urged to be antisymmetric about
$\omega_{4s,3s}$. Each pattern is composed of 26 bins which can
obtain values randomized in the [$-\pi/2,\pi/2$] interval.
On Fig.~\ref{fig_2} panel (a) presented a two-dimensional histogram of the
measured population: x-axis represents the population of 4s state,
y-axis represents the population of 7p state. The color code
and the topological curves represent the occurrence of phase patterns
with the corresponding signal. All the
results presented here are given relative to the absorption induced
by the transform-limited (TL) pulse.
For convenience, we introduce the TL-normalized absorption measures
$\widetilde{P}_{f_1} = P_{f_1} / P_{f_1,TL}$ and
$\widetilde{P}_{f_2} = P_{f_2} / P_{f_2,TL}$. The three regions,
corresponding to the constant, antisymmetric, and random phase
patterns are clearly distinguished on the histogram.
The region with the peak at ($\widetilde{P}_{4s}$=1,
$\widetilde{P}_{7p}$=0.5) correspond to the antisymmetric sequences,
encoded to the phase patterns, while the region with the peak at
($\widetilde{P}_{4s}$=0.25, $\widetilde{P}_{7p}$=0.25) correspond to
the random sequences. The region with the peak at
($\widetilde{P}_{4s}$=1, $\widetilde{P}_{7p}$=1) correspond to the
constant sequences with the populations identical to the TL pulse.
The broadening of that peak directly reflects the signal-to-noise
experimental ratio. As seen from the figure, the measured
population, corresponding to the sequence encoded to the phase
pattern of the excitation pulse can be used for the classification
of the sequence by its properties, based on the presented above truth
table.

On Fig.~\ref{fig_2} panels (b) and (c) we draw the projections of the two-dimensional
histogram of panel (a) on the 7p and 4s population axis. As seen from panel (c),
the 4s population signal clearly distinguish between the random sequences and the rest:
100$\%$ of the random sequences produce signal less then the corresponding TL pulse. However,
this signal do not distinguish between the constant and antisymmetric sequences. The latter is done by
analyzing the 7p population signal [see panel (b)].
For the detailed analysis of the distribution
we plot the cumulative sum function corresponding to the antisymmetric sequences in dashed black line and
obtain that 96$\%$ of the sequences produces 7p population signal less then the corresponding
TL pulse. It means that the experimental false identification of antisymmetric sequence as
a constant one or vise versa is below 4$\%$. This value can be improved
by improving the signal-to-noise ratio.

We have checked the sensitivity of the method by shifting the phase
pattern center of symmetry by one pixel apart from $\omega_{4s,3s}$.
The corresponding results are shown on Fig.~\ref{fig_2} panel (b).
The distribution of the antisymmetric patterns is moved from the region
of $\widetilde{P}_{4s}=1$ to the region corresponding to random patterns.
The reason for the high sensitivity of the signal to the exact location of the
inversion point is in the very narrow nature of the atomic resonance.
While the phase pattern is antisymmetric about $\omega_{4s,3s}/2$ it induces completely
constructive interference of all the photoinduced pathways to the $\left|4s\right>$ state.
When the inversion point is shifted by $\Delta$, the maximally constructive
interference would be induced for the state corresponding to the frequency $\omega_{4s,3s}+\Delta$.
In our experiment we have shifted the pattern by $\Delta=\delta\omega_{shaping}$=2.05cm$^{-1}$.
This shift was sufficient to reduce the constructive interference resulted in the reduced signal measured
for that case.

To summarize, we have demonstrated an efficient solution to the unknown sequence
classification problem. The classification is into three possible groups: a constant
sequence, a sequence that is antisymmetric about specific point, and a random sequence. The sequences
are encoded into the phases of the broadband femtosecond pulses. In the spirit of
the analog quantum computation, the photoinduced dynamics of the excited atomic sodium is used to
perform the computational task. The solution to the problem is obtained within
the measurement of the two excitation channels irrespectively of the sequence length.

This research was supported by The Israel Science Foundation (grant
No. 127/02), by The James Franck Program in Laser Matter
Interaction, and by The Technion's Fund for The Promotion of
Research.

\newpage

\newpage

\end{document}